\documentclass[preprint2]{aastex62}

\usepackage{natbib}

\usepackage{amsmath}
\usepackage{listings}
\usepackage{color}
\usepackage{float}
\usepackage{xspace} 
\usepackage{longtable}

\newcommand{\kepler}{\textsl{Kepler}\xspace}

\newcommand{\logrprime}{$\log R^\prime_\mathrm{HK}$\xspace}
\newcommand{\teff}{\ensuremath{T_{\mbox{\scriptsize eff}}}}
\newcommand{\logg}{\ensuremath{\log g}}
\newcommand{\vsini}{\ensuremath{v \sin i}}
\newcommand{\kms}{\ensuremath{\mbox{km s}^{-1}}}

\begin{document}

\title{Are Starspots and Plages Co-Located on Active G and K Stars?}

\author[0000-0003-2528-3409]{Brett M. Morris}
\affiliation{Astronomy Department, University of Washington, Seattle, WA 98195, USA}

\author[0000-0002-2792-134X]{Jason L. Curtis}
\altaffiliation{NSF Astronomy and Astrophysics Postdoctoral Fellow}
\affiliation{Department of Astronomy, Columbia University, 550 West 120th Street, New York, NY 10027, USA}

\author[0000-0001-7371-2832]{Stephanie T. Douglas}
\altaffiliation{NSF Astronomy and Astrophysics Postdoctoral Fellow}
\affiliation{Harvard-Smithsonian Center for Astrophysics, 60 Garden St, Cambridge, MA 02138}

\author[0000-0002-6629-4182]{Suzanne L. Hawley}
\affiliation{Astronomy Department, University of Washington, Seattle, WA 98195, USA}

\author[0000-0001-7077-3664]{Marcel A.~Ag{\"u}eros}
\affiliation{Department of Astronomy, Columbia University, 550 West 120th Street, New York, NY 10027, USA}

\author[0000-0002-5662-9604]{Monica G. Bobra}
\affiliation{W. W. Hansen Experimental Physics Laboratory, Stanford University, Stanford, CA 94305, USA}

\author[0000-0002-0802-9145]{Eric Agol}
\affiliation{Astronomy Department, University of Washington, Seattle, WA 98195, USA}
\altaffiliation{Guggenheim Fellow}

\email{bmmorris@uw.edu}

\begin{abstract}
We explore the connection between starspots and plages of three main-sequence stars by studying the chromospheric and photospheric activity over several rotation periods. We present simultaneous photometry and high-resolution ($R\sim 31,500$) spectroscopy of KIC 9652680, a young, superflare-producing G1 star with a rotation period of 1.4 days. Its \kepler light curve  shows rotational modulation consistent with a bright hemisphere followed by a relatively dark hemisphere, generating photometric variability with a semi-amplitude of 4\%. We find that KIC 9652680 is darkest when its $S$-index of \ion{Ca}{2} H \& K emission is at its maximum. We interpret this anti-correlation between flux and $S$ to indicate that dark starspots in the photosphere are co-located with the bright plages in the chromosphere, as they are on the Sun. Moving to lower masses and slower rotators, we present {\it K2} observations with simultaneous spectroscopy of EPIC 211928486 (K5V) and EPIC 211966629 (K4V), two active stars in the 650 Myr-old open cluster Praesepe. The {\it K2} photometry reveals that both stars have rotation periods of 11.7 days; while their flux varies by 1 and 2\% respectively, their \ion{Ca}{2} H \& K $S$-indices seem to hold relatively constant as a function of rotational phase. This suggests that extended chromospheric networks of plages are not concentrated into regions of emission centered on the starspots that drive rotational modulation, unlike KIC 9652680. We also note that the \ion{Ca}{2} emission of EPIC 211928486 dipped and recovered suddenly over the duration of one rotation, suggesting that the evolution timescale of plages may be of order the rotation period. 
\end{abstract}

\object{EPIC 211928486, EPIC 211966629, KIC 9652680}


\section{Introduction}

Sunspots are regions of the solar photosphere where convection is suppressed by strong magnetic fields \citep{Spruit1977, Solanki2003}. These spots typically account for a small portion of the  photosphere, covering only a fraction of a percent of the solar hemisphere \citep{Howard1984}.  Approximately a thousand kilometers above the photosphere lies the chromosphere. Observing the chromosphere in the cores of the Ca II H or K lines reveals regions of bright emission known as plages. Plages show a latitude dependence similar to sunspots, and the largest solar plages are typically spatially associated with dark sunspots. However, smaller plages often appear without cospatial sunspots, although at scales larger than 4 arcsec darker photospheric features, due to flux emergence, are observed to be cospatial with Ca II enhanced sites \citep{Hall2008, Guglielmino2010, Mandal2017}.

For distant stars, we are not privy to the same spatially resolved view. However, we can probe the connection between plages and starspots with time resolved spectroscopy and precision photometry. Plages are bright in the cores of the \ion{Ca}{2} H \& K features, while the broadband \kepler flux of a star, for example, is diminished when starspots face the observer \citep{Walkowicz2013,McQuillan2013,McQuillan2014,Douglas2014,Douglas2016,Douglas2017}. In this work, we present a study of chromospheric \ion{Ca}{2} H \& K emission and photospheric starspot absorption on the rotation period timescale for three stars to address the question: \textit{are plages and starspots co-located on active G and K stars}? 

Synoptic observations of chromospheric emission in \ion{Ca}{2} H \& K among many stars suggest that the emission variations of older, Sun-like stars increase with the stellar photometric variation; whereas young stars tend to become fainter overall as emission from plages increases \citep{Radick1998, Radick2018}. These observations are based on long-term monitoring campaigns of many stars with multi-epoch photometry and spectroscopy spread over years, and represent the effect of the activity cycle on the total irradiance and chromosopheric emission of each star. 

\begin{figure}
    \centering
    \includegraphics[scale=0.65]{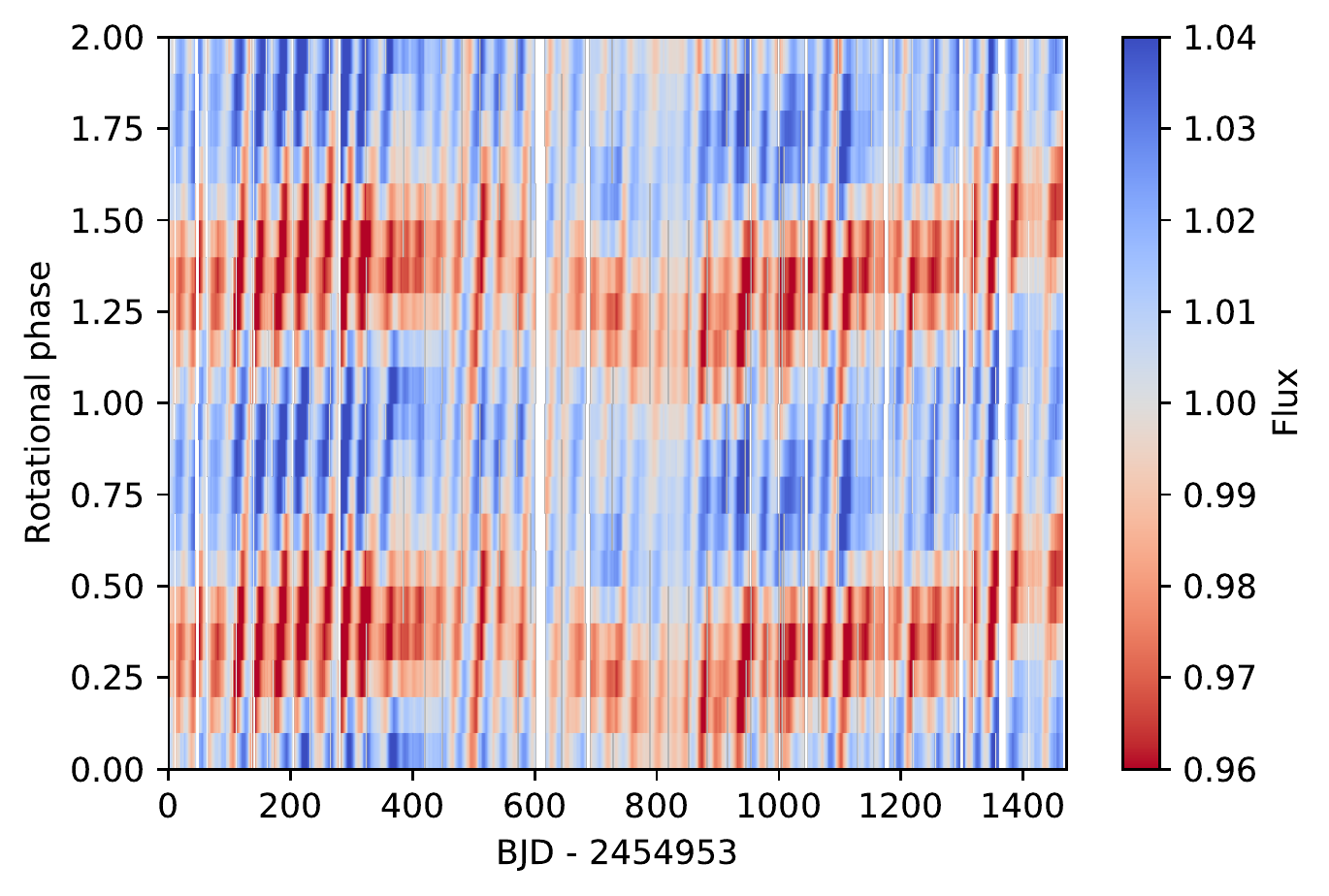}
    \caption{Rotational map of KIC 9652680. Each column shows one rotation (repeated twice for clarity, as in \citep{Davenport2015}) of the star, where color corresponds to relative flux of the star as a function of rotational phase. Each row is therefore a time-series of fluxes at one longitude, averaged over all latitudes, spanning the duration of the \kepler mission. The map shows one predominantly bright hemisphere, and the other darker, which remains roughly stable over the four year \kepler mission.}
    \label{fig:rotation}
\end{figure}

We collected simultaneous ground-based photometry and spectroscopy of young G star KIC 9652680 over the course of several rotations. This star is remarkable for having 8\% amplitude flux variations in the \kepler band -- see Figure~\ref{fig:rotation} -- and for producing a ``super-flare'' \citep{Notsu2015a}. The largest observed flare of KIC 9652680 reached $10^{34}$ erg, or roughly 100$\times$ more energy than the Carrington event \citep{Carrington1859, Neuhauser2014, Notsu2015b}. We present a similar analysis of the chromospheric and photospheric activity on this possible proxy for the young Sun. 

We also present simultaneous {\it K2} photometry and spectroscopy of two mid-K dwarfs in the young (650 Myr), nearby Praesepe cluster (M44). Confined to the ecliptic, the K2 mission provided precision photometry of Praesepe members in Campaigns 5 and 16 \citep{k2}. Campaign 16 was remarkable for being a ``forward-looking'' campaign, meaning that simultaneous observations were possible from ground-based observatories. We took advantage of that opportunity and collected more than 30 hours of high resolution spectra of EPIC 211928486 and EPIC 211966629 throughout the campaign, allowing us to study the connection between chromospheric and photospheric activity.  These two Praesepe members may be young proxies for the well-studied mid-K dwarf HAT-P-11 \citep{Deming2011, Sanchis-Ojeda2011, Morris2017a, Morris2017b, Morris2018d}.

We describe our spectroscopy of all targets in Section~\ref{sec:spectroscopy}. We measure the rotation periods and flux-activity correlation of the young G star in Section~\ref{sec:gstar}, and do the same for two K-dwarfs in Section~\ref{sec:k2}. We discuss the results in Section~\ref{sec:discussion} and conclude with Section~\ref{sec:conclusion}.

\section{APO/ARCES Spectroscopy} \label{sec:spectroscopy}

\begin{deluxetable*}{lccc}
\tablecaption{Stellar parameters\label{tab:stellarprops}}
\tablehead{ & \colhead{KIC 9652680} & \colhead{EPIC 211928486} & \colhead{EPIC 211966629}}
\startdata
$T_\mathrm{eff}$ [K] & $5900 \pm 100$ & $4408\pm 100$ & $4484\pm 100$ \\
$\log g$ & 4.47 & 4.66 & 4.66 \\
$[$Fe/H$]$ & $+0.21$ & $+0.12$ & $+0.12$ \\
$v \sin  i$ [km s$^{-1}$] & 39 & $2\pm2$ & $3\pm2$ \\
$P_\mathrm{rot}$ [d] & $1.41 \pm 0.01$ & $11.8 \pm 0.1$ & $11.6 \pm 0.2$ \\
$\log R^\prime_\mathrm{HK}$ & $-3.92 \pm 0.03$ & $-4.16 \pm 0.03$ & $-4.11 \pm 0.02$  \\
$f_{S, \rm min}$ & 0.26 & 0.041--0.046 &  0.061--0.083
\enddata
\end{deluxetable*}

The ARC Echelle Spectrograph (ARCES) on the ARC 3.5 m Telescope at Apache Point Observatory is an $R\approx31,000$ cross-dispersed spectrograph.  This analysis closely follows the procedure of \citet{Morris2017b} and makes use of its software \citep{aesop}. After stacking all exposures on a given night, we reduce the raw ARCES spectra with \texttt{IRAF} methods to subtract biases, remove cosmic rays, normalize by the flat field, and do the wavelength calibration with exposures of a thorium-argon lamp.\footnote{An ARCES data reduction manual by J.~Thorburn is available at \url{http://astronomy.nmsu.edu:8000/apo-wiki/attachment/wiki/ARCES/Thorburn_ARCES_manual.pdf}} We fit the spectrum of an early-type star with a high-order polynomial to measure the blaze function, and we divide the spectra of each star by the polynomial fit to normalize each spectral order.

Next, the normalized spectra must be shifted in wavelength into the rest-frame by removing their radial velocities (RVs). We remove the radial velocity by maximizing the cross-correlation of the ARCES spectra with PHOENIX model atmosphere spectra \citep{Husser2013}.

\section{Young G Star: KIC 9652680} \label{sec:gstar}
\subsection{Stellar Properties}
KIC 9652680 is a rapidly-rotating, 
chromospherically active, G-type star. 
Rapid rotation is often interpreted as a sign of youth, 
but can also be caused by a tidally-interacting companion (e.g., a RS CVn type binary, \citealt{Eaton1979})
or as a result of blue straggler formation (e.g., mass accretion \citealt{bluestraggler}), 
though these scenarios require the presence of a stellar companion. 
According to the 
second data release of the European Space Agency's \textit{Gaia} Mission \citep[DR2][]{GaiaDR2}, 
this star has RV error 3.42 \kms\ from 12 observations
\citep{DR2RV1, DR2RV2}; 
we measured RVs from our 13 ARCES observations and found a 
standard deviation of $\approx$1 \kms. These low RV values 
are consistent with KIC 9652680 being a single, rapidly rotating star, 
which is suggestive of youth. 

Its age is further corroborated by the presence of strong Li absorption at 6707.8 \AA, where $A$(Li) = 3.39 dex \citep{Honda2015}, which is unexpected in the alternative scenarios.
Considering its rotation and activity via the gyrochronology and activity--age relations 
and cluster activity data presented in \citet{Mamajek2008}, 
together with the lithium data for the Pleiades provided by  \citet{Bouvier2018}, 
KIC 9652680 appears to be at least as young 
as the Pleiades, or $\lesssim$110-125 Myr \citep{DR2HRD, Stauffer1998}. 

Its rapid rotation, and resulting $\vsini \approx 39$ \kms,
complicates our spectroscopic characterization of KIC 9652680.
However, if it is in fact a young star 
on or near the zero-age main-sequence, 
then its absolute magnitude or luminosity is 
actually a powerful indicator of metallicity.
We therefore compare KIC 9652680 to other young stars from \textit{Gaia} DR2 in order to constrain its metallicity. 

\begin{figure}
    \centering
    \includegraphics[scale=0.55]{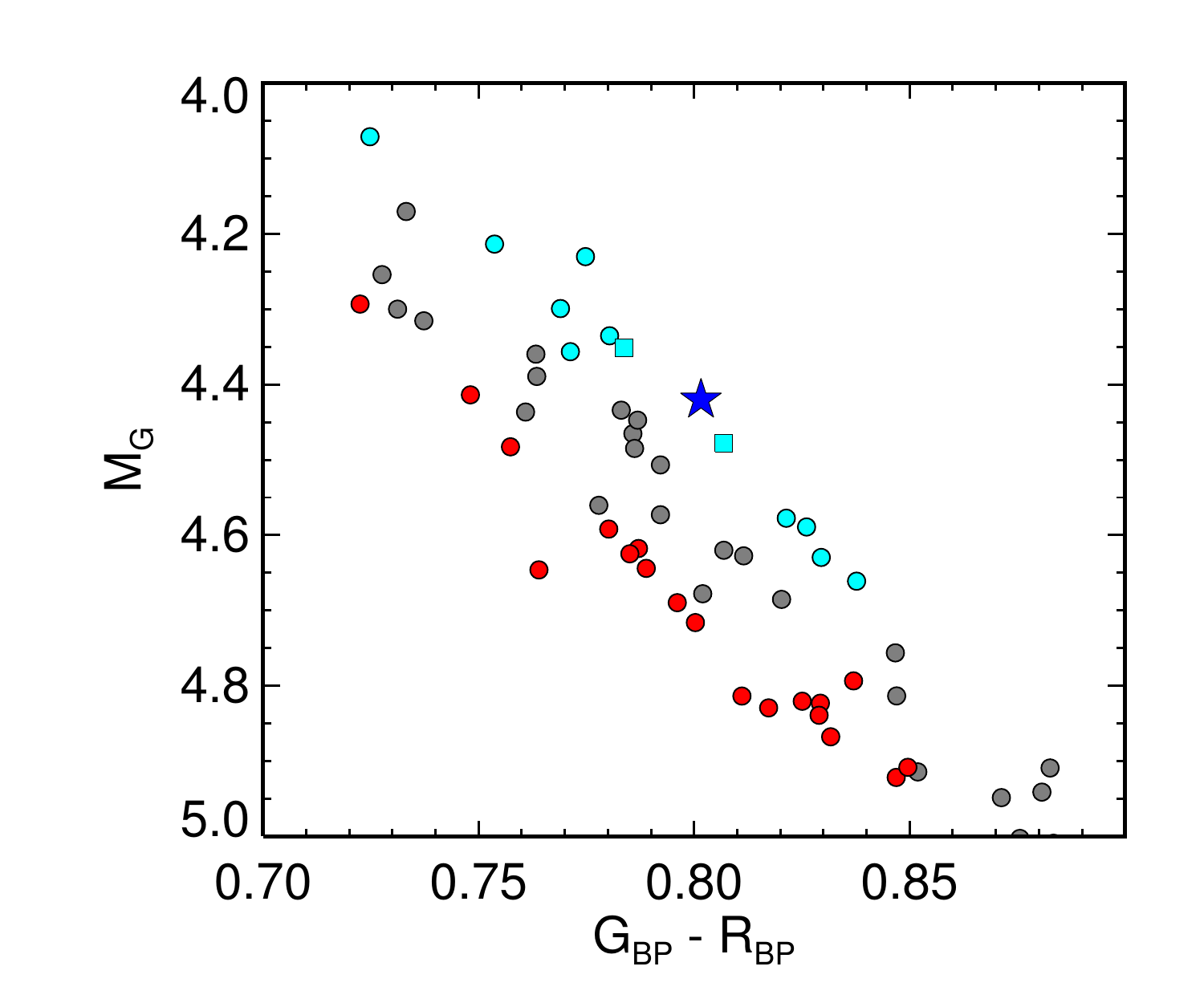}
    \caption{\textit{Gaia} DR2 CMD of KIC 9652680 (blue star) and chromospherically active (i.e., young, with \logrprime $< -4.7$ dex) stars observed with Keck/HIRES by the California Planet Survey and characterized with SME 
    by \citet{Brewer2016}. 
    Stars are color-coded according to their 
    metallicity, where red points have [Fe/H] $< 0.0$ dex, 
    cyan points have [Fe/H] $> +0.16$ dex,
    and gray points have intermediate metallicities.
    Assuming all stars shown are photometrically single 
    and have evolved in isolation as single stars,
    then the placement of KIC 9652680 in this CMD is indicative of its 
    metallicity, as any such star with \logrprime $\approx -4.0$ dex
    must be very young and on or near the zero-age main-sequence.
    The stars HD 222986 and HD 101847, marked with cyan-shaded squares, 
    bracket KIC 9652680 in this diagram, indicating that it likely has 
    properties intermediate to their values of \teff\ = 5858 and 5948 K 
    and [Fe/H] = +0.26 and +0.23 dex, respectively, with \logg\ = 4.47 dex.}
    \label{fig:CMDkic}
\end{figure}

Figure \ref{fig:CMDkic} shows 
the \textit{Gaia} DR2 color--magnitude diagram (CMD; $G_{\rm BP} - R_{\rm RP}$ vs $M_G$)
for chromospherically active (i.e., relatively young stars, 
with \logrprime $< -4.7$ dex corresponding to ages under 2 Gyr),
solar-type stars observed with Keck/HIRES 
by the California Planet Survey (CPS) and characterized with 
Spectroscopy Made Easy \citep[SME;][]{sme, Valenti2005, Piskunov2017}
by \citet{Brewer2016} with the procedures outlined in \citet{Brewer2015}.

We excluded stars with $\logg < 4.3$ dex to ensure 
that we are focusing on un-evolved dwarf stars, 
and included stars with HIRES spectra with $S/N > 70$ per pixel
to select those with precise spectroscopic properties,
and focused on stars with precise DR2 photometry 
($\sigma < 0.05$ mag)
with parallax-inverted distances within 200 pc 
to minimize the effect of interstellar reddening and extinction 
on the CMD.
The majority of these stars are presumably single, 
and on or near the main-sequence. 
The stars in Figure~\ref{fig:CMDkic} are color-coded according to their spectroscopic metallicities; as expected, metallicity explains the majority of the 
apparent scatter in this CMD, 
and the higher metallicity stars follow a 
more luminous track in the CMD. 
KIC 9652680 is found at the top of
the main-sequence envelope 
among a string of 8 stars with the following properties 
(median $\pm$ standard deviation): 
[Fe/H] = +0.205 $\pm$ 0.024 dex, 
\logg\ =  4.475 $\pm$ 0.023 dex, 
and 
\teff\ =  5844 $\pm$ 90 K. 
Fitting a quadratic function to 
$(G_{\rm BP} - G_{\rm RP})$ vs. \teff\ for this sequence, 
KIC 9652680 has $\teff = 5900 \pm 20$ K,
where the uncertainty is calculated as the 
rms of the spectroscopic values about our 
quadratic color--\teff\ relation.

The two stars with $(G_{\rm BP} - G_{\rm RP})$ closest to KIC 9652680
(marked as cyan-colored squares in Figure~\ref{fig:CMDkic}),
and which happen to bracket it, have the following spectroscopic properties: 
HD 222986 has \teff\ = 5858 K, \logg\ = 4.47 dex, 
[Fe/H] = +0.26 dex, \vsini\ = 5.8 \kms, 
and \logrprime = $-4.35$ dex; 
HD 101847 has \teff\ = 5948 K, \logg\ = 4.47 dex, 
[Fe/H] = +0.23 dex, \vsini\ = 5.6 \kms, 
and \logrprime = $-4.43$ dex; 
The \vsini\ values are much lower than KIC 9652680, 
which allowed \citet{Brewer2016} to measure 
precise stellar properties from the HIRES spectra.
According to the \citet{Mamajek2008} 
activity--rotation--age relation, 
their chromospheric emission indices 
imply ages of 105 and 227 Myr, 
which is young enough to assume negligible 
main-sequence evolution through the CMD for these stars.
From this analysis, 
we find the following properties 
for KIC 9652680 from our 
\textit{Gaia}--CPS analysis: 
\teff = 5900 K (based on the color--\teff\ relation), 
\logg = 4.47 dex (based on these two stars), and 
[Fe/H] = +0.21 dex (based on the median for the 8 star sequence).

Next, we analyzed our 13 ARCES observations of KIC~9652680 with SME version 522
using the \citet{Valenti2005} strategy, 
which focuses on 
the 5164--5190 \AA\ region centered on the 
pressure-sensitive \ion{Mg}{1} $b$ triplet 
and seven additional segments spanning 6000--6180 \AA. 
We 
ran SME with \teff, \logg, the global metallicity parameter [M/H], 
and \vsini\ as free parameters; 
this yielded 
$\teff = 6022 \pm 99$ K, 
$\logg = 4.40 \pm 0.1$ dex, 
[M/H] = $+0.17 \pm 0.08$ dex, and 
$\vsini = 39 \pm 0.5$ \kms. 

The \teff\ from our ARCES spectra is warmer than our
\textit{Gaia}--CPS value by 122 K 
(1.2 $\sigma$ using the spectroscopic \teff\ dispersion for the uncertainty), 
and this discrepancy and the high dispersion in fitted parameters 
is likely 
due to the challenges 
of analyzing spectra of 
rapidly rotating stars, 
especially with such a restricted wavelength range 
compared to the full optical spectrum observed with 
ARCES.
For this work, 
we will take the results from our 
\textit{Gaia}--CPS analysis as our 
final adopted stellar properties for KIC 9652680, 
and assign an uncertainty to \teff\ of 100 K to 
reflect the discrepancy with the SME result. 
The final values are listed in Table~\ref{tab:stellarprops}.


\subsection{Rotational modulation}

\begin{figure*}
    \centering
    \includegraphics[scale=0.9]{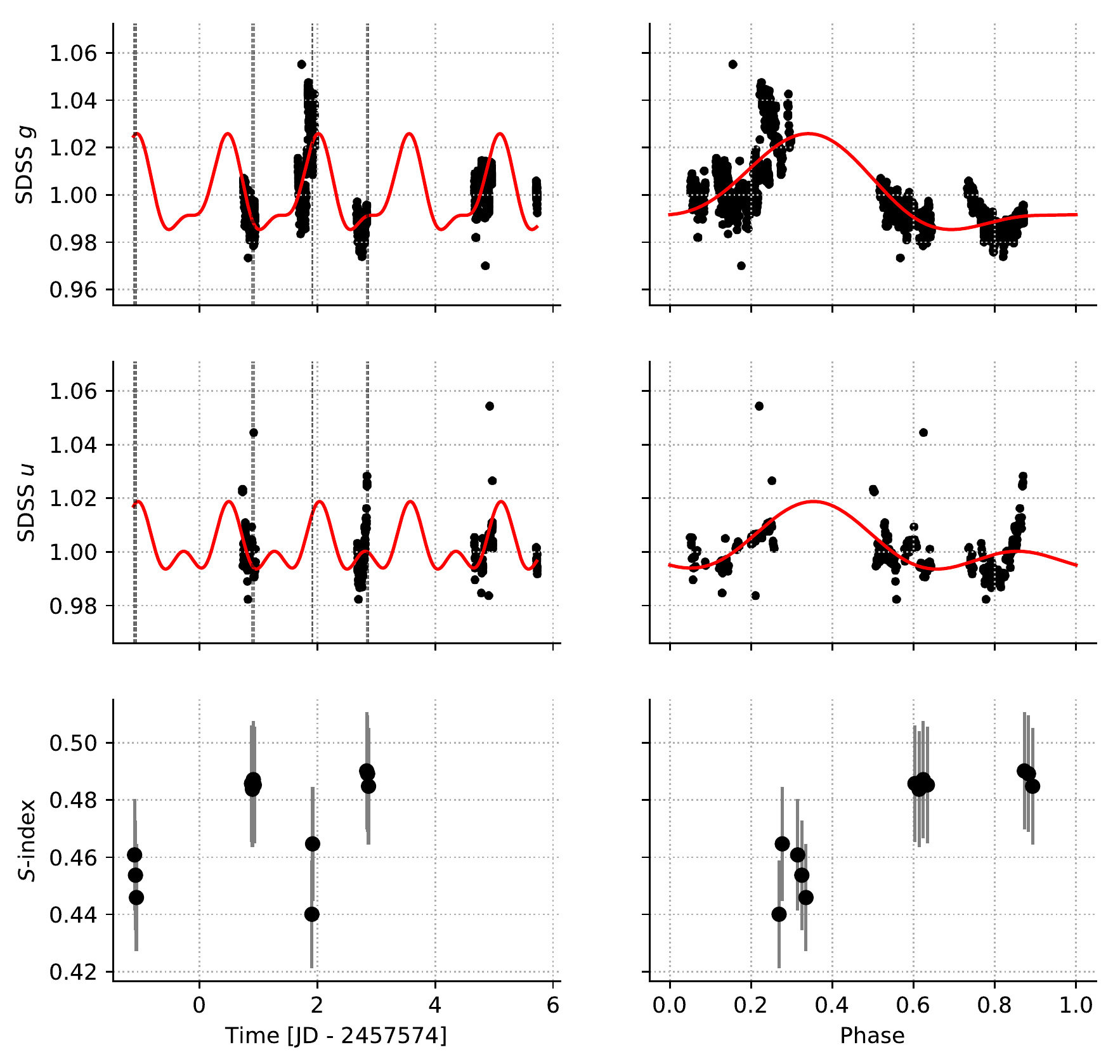}
    \caption{ARCSAT and ARCES monitoring campaign of KIC 9652680 to constrain the rotational phase at each spectroscopic measurement. The left column shows the time-series photometry in SDSS $g$ and $u$ bands, along with the \ion{Ca}{2} H \& K $S$-index measured as a function of time. The red curve shows the best-fit multiband Lomb-Scargle periodic model to the photometry; the vertical dotted lines mark the times of the spectroscopic observations. The right column shows the same information phased to the rotation period of the star.}
    \label{fig:arcsat}
\end{figure*}

The \kepler PDCSAP light curve of the G1V star KIC 9652680 shows rotational modulation consistent with a bright hemisphere followed by a relatively dark hemisphere, generating photometric variability with semi-amplitude 4\%, see Figure~\ref{fig:rotation}. We compute the rotation period and its uncertainty from the \kepler observations by taking the mean and difference between the Lomb-Scargle periodogram and the first peak in the autocorrelation function, finding $P_\mathrm{rot} = 1.41 \pm 0.01$ d.

We observed KIC 9652680 over five nights in the SDSS $u$ and $g$ bands with FlareCam on the ARCSAT 0.5 m telescope at Apache Point Observatory; see Figure~\ref{fig:arcsat}. We computed differential photometry with three neighboring comparison stars -- our ARCSAT photometry pipeline is open source and available online\footnote{\url{https://github.com/bmorris3/arcsat_monitoring}}. 

We model the flux variations in both bands simultaneously with the multiband Lomb-Scargle (MLS) technique of  \citet{VanderPlas2015} with two sinusoidal terms (red curves). MLS estimates the relative flux of the star in each band assuming the periodic signals have the same phase but different amplitudes in each band. The flux variations that we observed with ARCSAT show similar rotational modulation amplitude in the SDSS $g$ band compared to the \kepler semi-amplitude of 4\%. The MLS prediction allows us to interpolate between observed fluxes to estimate the flux of the star at the epochs of each ARCES spectrum, to search for correlations between the $S$-index and broadband flux of the host star. 

\subsection{Minimum spot covering fractions via flux deficit}

We can compute the minimum spot covering fractions required to produce the observed rotational modulation from the flux deficit, via
\begin{equation}
f_\mathrm{S, min} = \frac{1 - \min{\mathcal{F}}}{1-c}
\end{equation}
where $\mathcal{F}$ is the stellar flux, and $c$ is the spot contrast (where $c=0$ is a spot with the same intensity as the photosphere and $c=1$ is a spot that is perfectly dark). Assuming the same average contrast as sunspots, $c=0.3$ \citep[see e.g.][]{Morris2017a}, we compute spot covering fraction from the \kepler light curve, and find $f_{S, \min} \gtrsim 0.17$. This large spot covering fraction is roughly consistent with observations of young G stars like EK Draconis, for example, which has $f_S = 0.25-0.40$ according to TiO molecular band absorption measurements by \citet{ONeal2004}.

\subsection{Flux--$S$-index correlation}

\begin{figure}
\centering
    \includegraphics[scale=0.7]{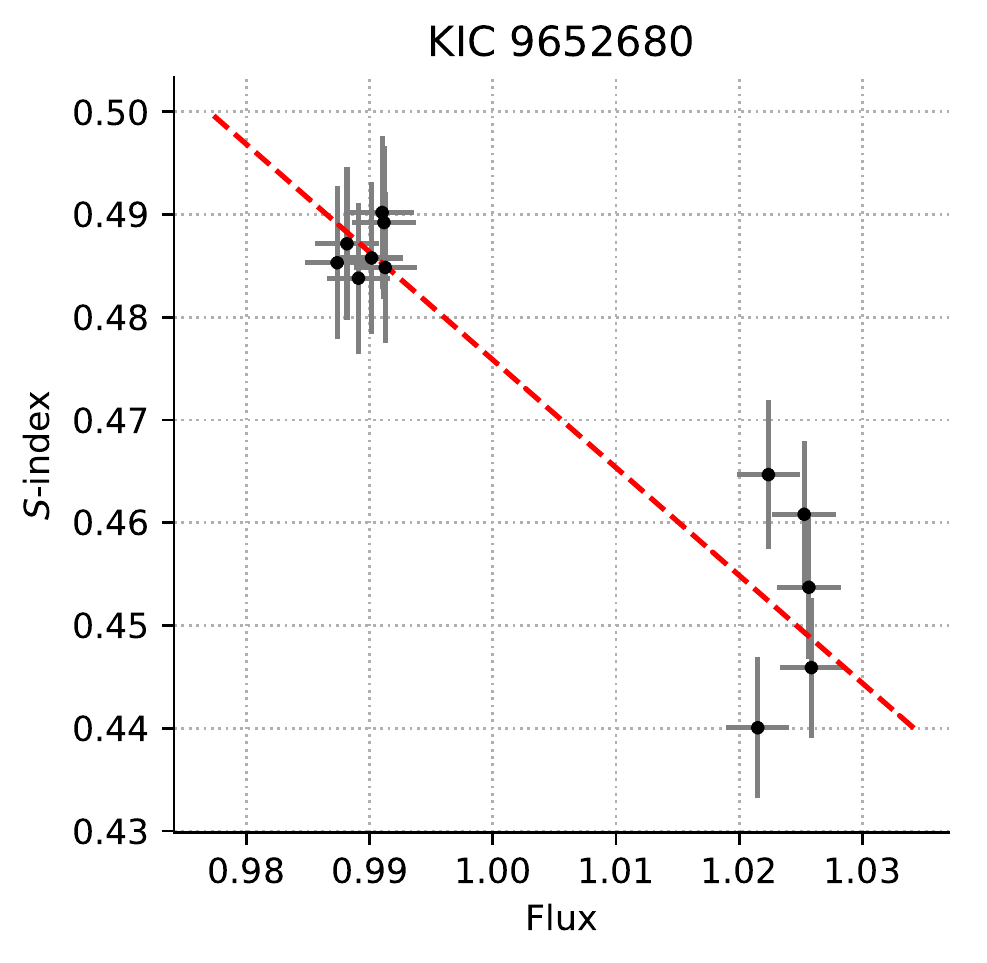}
    \caption{$S$-index of KIC 9652680 as a function of the stellar flux in the SDSS $g$ band. The anti-correlation may suggest that a portion of the chromospherically active region, which produces bright emission in \ion{Ca}{2} H \& K, is co-located with the photospheric dark starspot, which decreases the stellar flux.}
    \label{fig:sind_corr_kic}
\end{figure}

Figure~\ref{fig:sind_corr_kic} shows an anti-correlation between the photosphere flux in SDSS $g$ and the chromospheric emission from KIC 9652680. We interpret this anti-correlation as evidence for spatial association between regions of chromospheric activity and the starspots in the photosphere below, as is observed on the Sun. As the spots and their associated plages rotate into view, the $S$-index increases and SDSS $g$ flux decreases in sync.

We note that the minimum in KIC 9652680's observed $S$-index does not correspond to the minimum possible $S$-index for a star of KIC 9652680's color ($S\gtrsim 0.15$, \citealt[see e.g.][]{Isaacson2010}). We still observe significant chromospheric emission when the dark spot which dominates the rotational modulation is out of view -- thus the network of plages must extend around all rotational phases of the star, while a significant component of the chromospheric emission is associated with the dark spot.

\section{Mid-K Dwarfs: EPIC 211928486 and EPIC 211966629} \label{sec:k2}
\subsection{Stellar Properties}
EPIC 211928486 and EPIC 211966629 
are members of the benchmark open cluster Praesepe (M44, NGC 2632)\footnote{According to 
\textit{Gaia} DR2, 
each star is co-moving with Praesepe. 
EPIC 211966629 (Gaia DR2 661335288963844736) 
actually has a discrepant parallax 
that places it 20 pc beyond the cluster center, 
which we assume is why it was not classified as a member by 
\citet{DR2HRD}; 
however, we note that this might be due to 
its non-zero astrometric noise excess, 
and we will consider it a real member of Praesepe 
based on its consistent 3D kinematics, 
rotation, and chromospheric activity.}
which
has an age of 620-790 Myr
\citep{DR2HRD, Gossage2018, Cummings2017, Choi2016, Brandt2015}, 
a metallicity of [Fe/H] = +0.12-0.16 dex 
\citep[although various results in the literature 
range from +0.038 to +0.27 dex; see Table 4 in][]{Cummings2017}, 
and suffers little if any interstellar reddening and extinction
\citep[$E(B-V) = 0.027$,][]{Taylor2006}. 
\citet{Allen1995} classified these
stars with low-resolution optical spectra 
as K5V and K4V, respectively.

More recently, each star was observed with the 
Large Sky Area Multi-Object Fiber Spectroscopic Telescope \citep[LAMOST;][]{Luo2015}
and stellar properties and RVs 
were measured and provided in the third data release. 
EPIC 211928486 and EPIC 211966629 were both classified as K5 dwarfs with the following properties: 
\teff\ = 4413 K and 4487 K, 
\logg\ = 4.69 dex and 4.49 dex, and
[Fe/H] = +0.08 dex and $-0.03$ dex.

Each star appears to be effectively single. 
First, 
their \textit{Gaia} DR2 RVs have small errors 
($\sigma \lesssim$1 \kms) 
and are consistent with the cluster's bulk velocity.
We also measured relative RVs between the two stars 
on seven separate nights (six in Dec 2017, 
one in Feb 2018), 
and found that they are constant 
to within 0.5 \kms.
Second, 
EPIC 211928486 is within 0.02 mag of,
and 
EPIC 211966629 is 0.10 mag brighter than,
the Praesepe single-star main-sequence, 
according to the 
\textit{Gaia} DR2 CMD for Praesepe 
using the membership from \citet{DR2HRD}.
We attribute this modest apparent magnitude enhancement 
to the cluster's depth 
(at 186 pc, a cluster with an angular size of 2.5 degrees 
has a physical radius of $\sim$8 pc; 
this translates to a $\pm$0.09 mag distance modulus effect); 
alternatively, 
there could be a $\sim$M3V companion, but 
it would likely be non-interacting due to the observed 
low RV dispersion.

While \citet{DR2prop} provided \teff\ values inferred from the 
\textit{Gaia} DR2 photometry, 
we opted to re-calibrate the DR2 \teff\ scale
using the benchmark sample of K and M dwarfs 
characterized with interferometry and 
bolometric fluxes by \citet{Boyajian2012} by 
fitting a quadratic function between 
the DR2 color $(G_{\rm BP} - G_{\rm RP})$ and \teff\ for these stars.
We adopted the \citet{Taylor2006} value of 
$E(B-V) = 0.027$ and used it to de-redden the DR2 photometry for EPIC 211928486 and EPIC 211966629, 
then infer \teff\ = 4408 K and 4484 K, respectively, 
whereas the DR2 catalog listed values of 4683 K and 4922 K.
Our values are much closer to the LAMOST DR3 values, 
being only 5 and 3 K cooler, 
whereas the DR2 catalog values are 
275 K and 438 K warmer than our values. 
Combining our results with LAMOST, 
we conclude that 
EPIC 211966629 is only 75 K warmer than EPIC 211928486. 

We also fit the APO/ARCES spectra for these stars with SME, 
despite the fact that the \citet{Valenti2005} strategy we followed 
is optimized for late-F to early-K dwarfs.
According to \citet{DR2HRD}, 
Praesepe's DR2 CMD can be approximately described with a PARSEC isochrone \citep{parsec, Marigo2017}
with an age of 700 Myr and [Fe/H] = +0.12 dex. 
We decided to ignore the \ion{Mg}{1} $b$ order (5164-5190 \AA) which is 
challenging to accurately normalize for mid-K dwarfs, 
then fixed the metallicity to [Fe/H] = +0.12 dex,
and fixed \logg\ to 4.65 dex based on this isochrone.
We fit the spectrum for \teff\ and \vsini, 
then combined the results from the multiple observations 
as the median and standard deviation, and report the following properties:
for EPIC 211928486,
$\teff = 4510 \pm 28$ K and 
$\vsini = 2 \pm 2$ \kms; 
for EPIC 211966629, 
$\teff = 4592 \pm 19$ K, and 
$\vsini = 3 \pm 2$ \kms,
where the \vsini\ error is primarily due 
to the uncertain spectral resolution at each epoch, 
as well as the macroturbulent broadening.
Here, we found that 
EPIC 211966629 is 82 K warmer than EPIC 211928486, 
which is nearly identical to the relative \teff\ 
we found from our photometric calibration and the LAMOST DR3 value, 
although the spectroscopic \teff\ values are systematically warmer 
than these other results by $\sim$100 K.
We set this offset as our \teff\ uncertainty, 
and adopt the photometric results as our final values.
For metallicity, 
we adopt [Fe/H] = $+0.12 \pm 0.05$ dex (a value commonly cited for the cluster, with uncertainty estimated to encompass most modern 
estimates of the cluster's metallicity).
For surface gravity, we set 
\logg\ = 4.66 dex,
taken from the 700 Myr PARSEC isochrone
with this metallicity at these temperatures.
These final values are listed in Table~\ref{tab:stellarprops}.


\begin{figure*}
    \centering
    \includegraphics[scale=0.6]{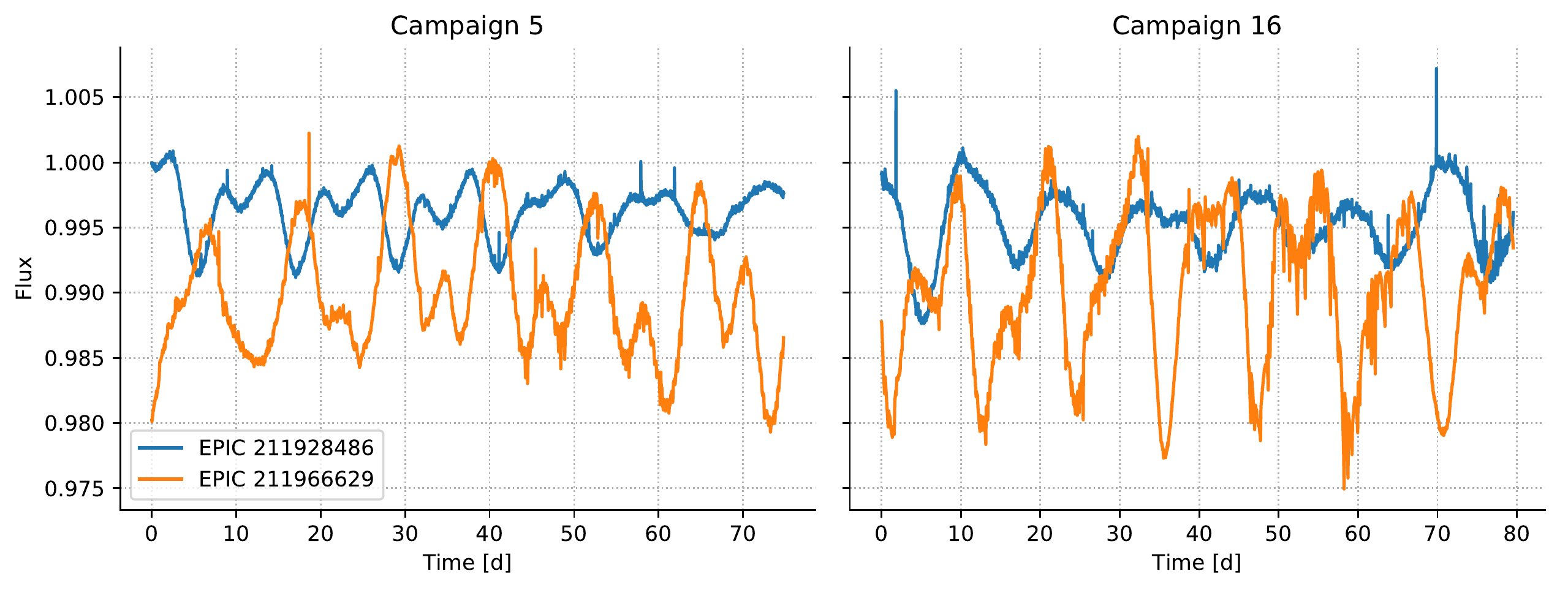}
    \caption{{\it K2} PDCSAP light curves for each target from Campaign 5 (left) and Campaign 16 (right).}
    \label{fig:lightcurves}
\end{figure*}

\subsection{Rotation Periods}

\begin{figure*}
    \centering
    \includegraphics[scale=0.7]{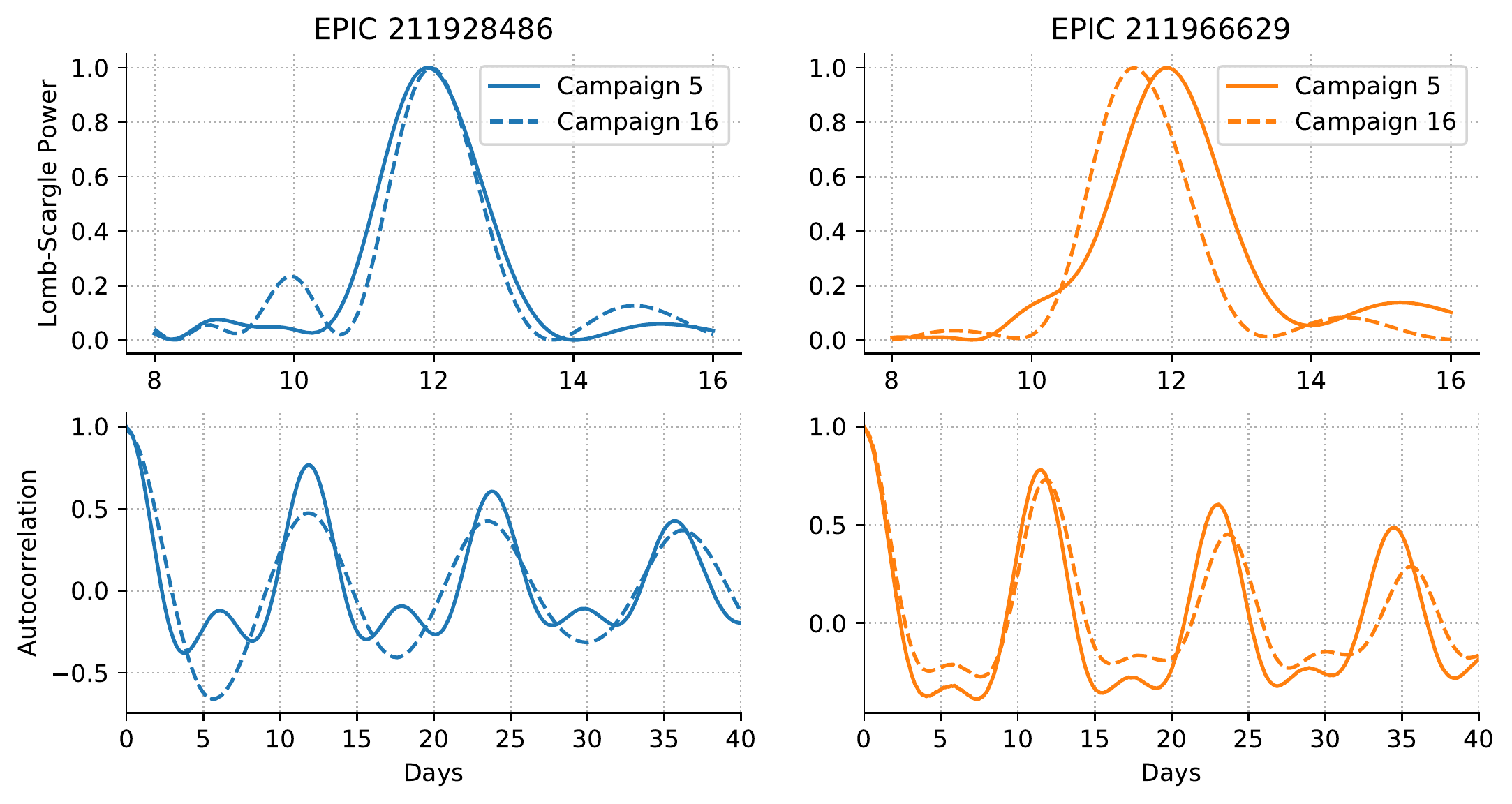}
    \caption{Lomb-Scargle periodogram and autocorrelation functions for the K2 photometry in Campaigns 5 and 16.}
    \label{fig:periodograms}
\end{figure*}

We use {\it K2} photometry processed with PDCSAP pipeline for Campaigns 5 and 16, see Figure~\ref{fig:lightcurves}. Both stars show rotational modulation consistent with 1-2 starspots. We measure the rotation periods for each star by taking the autocorrelation function and Lomb-Scargle periodogram for each target and each campaign; see Figure~\ref{fig:periodograms}. We take the mean and standard deviation of the four rotation period measurements (one LS and one ACF period for each campaign)  and find $P_\mathrm{rot} = 11.80 \pm 0.11$ days for EPIC 211928486, and $P_\mathrm{rot} = 11.68 \pm 0.22$ days for EPIC 211966629. 

The rotation periods of EPIC 211966629 measured in Campaigns 5 and 16 appear to be slightly different, though within the observational uncertainties. It is possible that the differing period measurement is astrophysical, and not an artifact of noise. If it is astrophysical, we may be measuring slightly different periods as starspots emerge at different latitudes on the star throughout the stellar activity cycle, giving a hint at differential rotation on this star. We refrain from making such a claim as previous work has made it clear that differential rotation is exceptionally hard to determine from rotational modulation \citep{Aigrain2015}.

\subsection{Minimum spot covering fractions via flux deficit}

We compute the flux deficit as in the previous section. The flux deficit indicates that both stars have similar spot covering fractions, between 1 and 4\%. Both stars are 100x more spotted than the Sun, and they appear to be similar to the K4 dwarf HAT-P-11 ($0 \leq f_S \leq 0.14$), which has a longer rotation period $P_\mathrm{rot} = 29$ d \citep{Morris2017a, Morris2018d}. If we assume that these stars are young proxies for HAT-P-11, it's interesting to note that while they have more chromospheric activity, their spot coverage is similar to their slower rotating counterpart HAT-P-11. 

\subsection{Lack of Flux--$S$-index correlation}

\begin{figure*}
    \centering
    \includegraphics[scale=0.8]{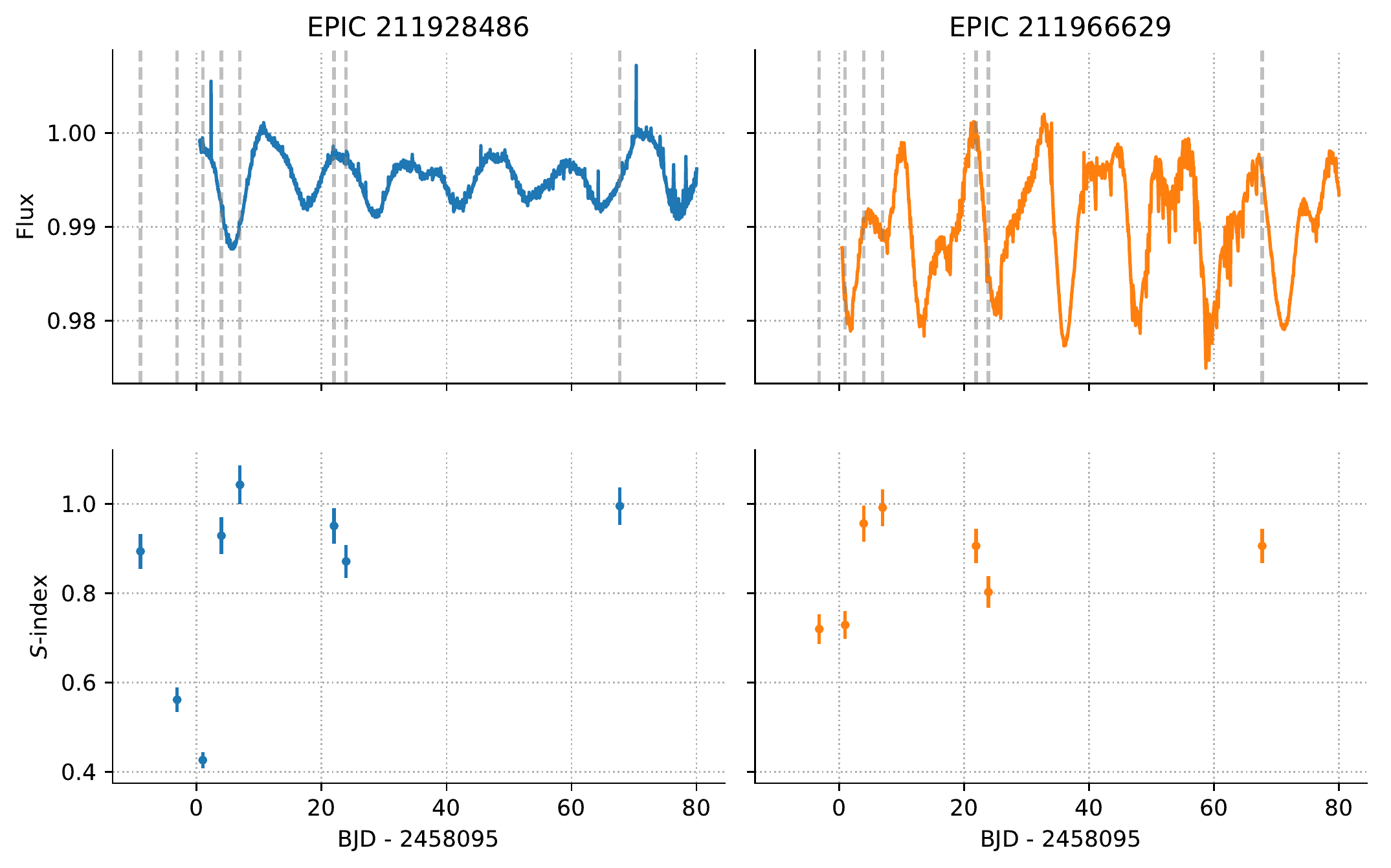}
    \caption{{\it K2} photometry with simultaneous ground-based \ion{Ca}{2} H \& K $S$-index monitoring with ARCES spectrograph on the ARC 3.5 m Telescope at APO. Vertical dashed lines on the upper row correspond to the times of spectra. }
    \label{fig:photometry_sindices}
\end{figure*}

\begin{figure}
    \centering
    \includegraphics[scale=0.8]{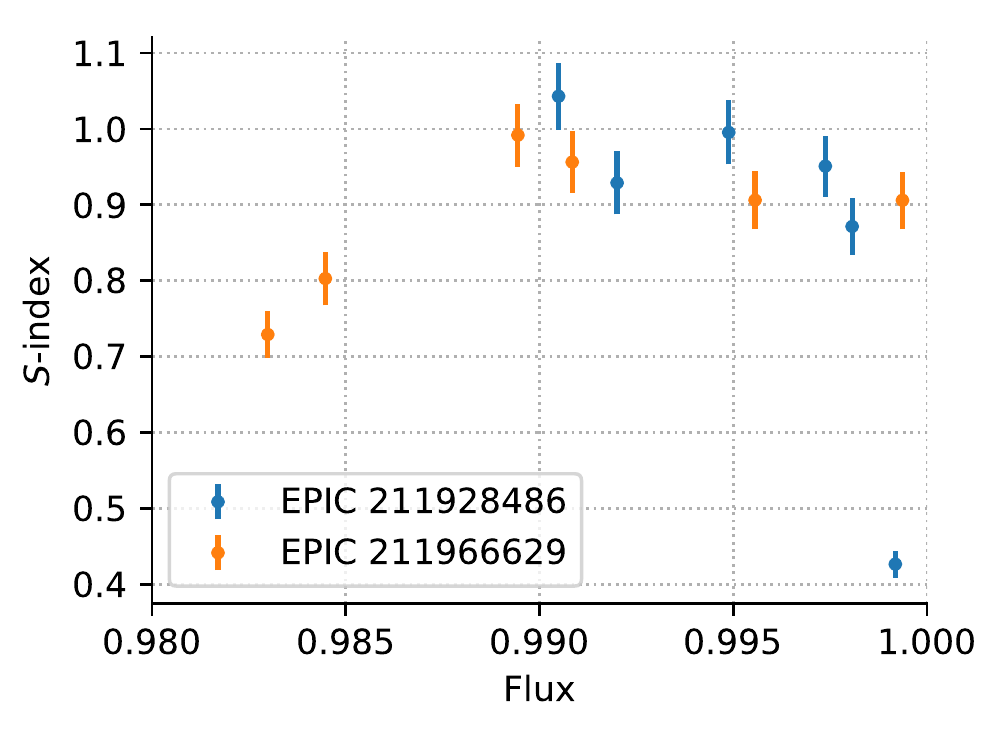}
    \caption{Comparing the $S$-index with the instantaneous stellar flux. EPIC 211928486 shows a weak anticorrelation between the chromospheric activity and the stellar flux, perhaps indicating that the dominant dark starspots are co-located with chromospherically active regions on the stellar surface. EPIC 211966629 shows no significant correlation between the flux and chromospheric activity, perhaps indicating that the chromospherically active regions are decoupled from the dominant starspots.}
    \label{fig:sind_corr}
\end{figure}

Figure~\ref{fig:photometry_sindices} shows the K2 photometry of each target (top panels), along with the $S$-indices that we measured with ARCES (bottom panels). Excluding the two spectra of EPIC 211928486 near BJD=2458095, the other 13 $S$-index measurements are more or less consistent with constant $S\approx 0.9$ for both stars, at all rotational phases. The lack of correlation between the observed stellar flux in the \kepler band and the $S$-index is demonstrated in Figure~\ref{fig:sind_corr}. This lack of correlation is interesting given that the corresponding $\log R^\prime_\mathrm{HK} \sim -4.1$ is generally considered ``very active'' \citep{Wright2004}, though the minimum spot coverage inferred from rotational modulation is only a few percent. 

One could interpret of the high $S$-index and small minimum spot coverage to imply that there are many plages without accompanying starspots, generating significant \ion{Ca}{2} H \& K emission at all rotational phases, but not contributing significantly to the rotational modulation in the \kepler band. This is frequently the case on the Sun. An alternative scenario is that the star is blanketed in spots which all generate \ion{Ca}{2} emission and photospheric absorption, and as each spot rotates into view, another rotates out of view, causing the relatively small changes in the observed \kepler flux ($\lesssim 2\%$). 

\subsection{Precipitous decline in $S$-index of EPIC 211928486}

\begin{figure}
    \centering
    \includegraphics{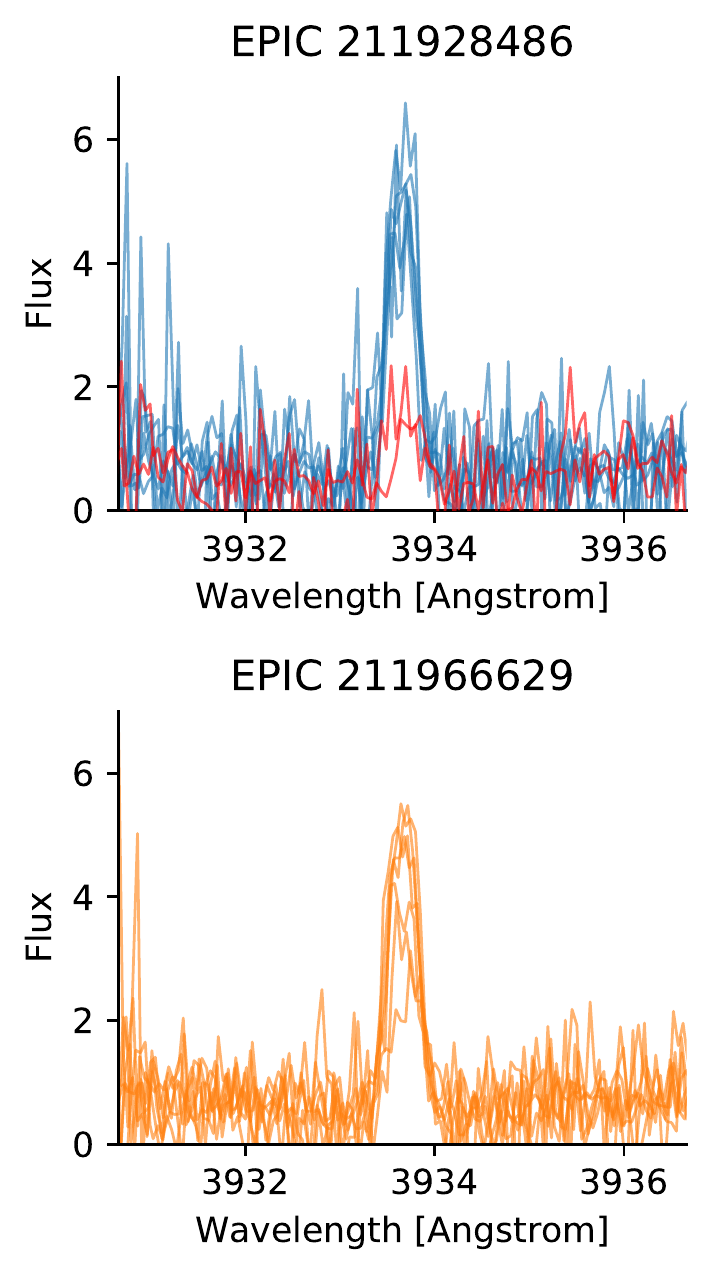}
    \caption{Close-in look at the \ion{Ca}{2} K line emission from each star, highlighting the anomalously low emission from two spectra near BJD=2458095 in red (see Figure~\ref{fig:sind_corr}).}
    \label{fig:kline}
\end{figure}

The $S$-indices of both stars are relatively constant with a significant exception. There is a 50\% decline in the emission in the cores of the \ion{Ca}{2} H \& K lines during two consecutive epochs; 
see Figures~\ref{fig:photometry_sindices} and \ref{fig:kline}. We have simultaneous K2 photometry for the second low-$S$ observation, but not the first (see Figure~\ref{fig:photometry_sindices}). We have inspected the spectra to verify that there were no additional sources of continuum emission, due to interloping nearby stars or solar contamination (reflected off of clouds or the sky, though these observations were not collected near twilight), which may appear to suppress the $S$-index. A flare could increase the continuum relative to the \ion{Ca}{2} emission \citep{Kowalski2013}, which would make the emission core appear weaker.
If a flare was responsible, however, we would also expect to see significant H$\alpha$ emission; we have verified that the H$\alpha$ line does not show any change during these epochs. 

As a result, it appears that one hemisphere had significantly less plages for a duration of about one rotation, and that the plages had recovered to their near-axisymmetric distribution by the next rotation. We are unaware of any observations in the solar literature that show a similar precipitous decline in chromospheric activity without an accompanying signature in the number of sunspots, indicating perhaps that this surprising phenomenon may be relevant to cooler or younger stars than the Sun. Continued observations of this star, preferably at higher cadence, would be valuable to constrain whether or not this phenomenon occurs regularly.

\section{Discussion} \label{sec:discussion}

A large sunspot in the solar photosphere is almost always spatially associated with a bright active region plages in co-temporal \ion{Ca}{2} K images of the chromosphere \citep{Stix1989, Mandal2017}. Thus we expected to find some degree of anti-correlation between the chromospheric emission and broadband flux for Sun-like stars with large spots/plages. We observed such an anti-correlation on the young G star KIC 9652680, perhaps indicative of large plages that accompany its most-spotted hemisphere. 

Similar association between plages and starspots has been observed many stars, including: 12 Oph (K1V) and 61 Cyg A (K5V) \citep{Dorren1982}, several stars in the famous sample of Olin Wilson's \citep{Radick1983}, several Hyades members \citep{Lockwood1984}, and the RS CVn variable UX Ari (K0IV) \citep{Gu2005}. 

We do not find a correlation between the broadband flux and chromospheric emission for the two mid-K stars EPIC 211928486 and EPIC 211966629. This may indicate that the plages are distributed axisymmetrically about the stellar surface, or that the spots that contribute to rotational modulation are too small to generate substantial chromospheric emission. From spot occultations during planetary transits of HAT-P-11 b, we know that the spots on mid-K stars can be as large as the largest spots on the Sun at solar maximum \citep{Morris2017a, Morris2018d}, and therefore one might expect plages to accompany those large spots. Thus we posit that the near-constant significant chromospheric emission of EPIC 211928486 and EPIC 211966629 is more likely due to an axisymmetric distribution of plages (and perhaps spots as well).  

The sudden decrease and recovery in the $S$-index of EPIC 211928486 may indicate that the evolution timescale of plages on these mid-K dwarfs is of order the rotation period. Large active regions on the Sun can last from days up to several rotations \citep{Solanki2003}. What we are observing is a curious \textit{lack} of activity lasting only a rotation, without an accompanying brightening of the star, as one might expect if the spots vanished, or dimming of the star, as one might expect if a planet or companion eclipsed the active regions. We invite observers to monitor the \ion{Ca}{2} emission of this star, for want of an explanation. 
We note that solar plages are observed at highest contrast with at the solar limb, and therefore their rotational signal on other stars will be slightly more complicated than the simplistic anticorrelation with total flux that we outline in this work. However, the brightening effect of faculae is small compared to the dimming due to starspots over the course of a full solar rotation. In broad optical bandpasses, the majority of the rotational modulation signal is produced by sunspots rather than faculae \citep{Shapiro2016}.

We also note that the observations presented here do not represent complete phase coverage for all stars. The photometric and spectroscopic observations are expensive and require large telescopes with dedicated observing programs, such as the one we have provided in this work. There are no other proxies for activity indices that we can draw upon for further analysis, other than the rotational modulation of the flux recorded by Kepler/FlareCam. 

In future work, it would be interesting to measure the full-disk $S$-index of the Sun as if it were a star, to determine: (1) whether or not the simple anticorrelation between broadband flux and $S$ indeed holds up for the Sun; and (2) whether or not the Sun exhibits dips significant dips in $S$ like EPIC 211928486.

\section{Conclusions} \label{sec:conclusion}

We have observed a young G star and two mid-K dwarfs with simultaneous photometry and spectroscopy to probe the connection between the distribution of plages and starspots on their surfaces as they rotate. We find that the young G star KIC 9652680 exhibits Sun-like behavior: when its most spotted hemisphere is facing us and its broadband \kepler flux is minimal, we observe enhanced chromospheric emission. 

The same is not true for the pair of K dwarfs EPIC 211928486 and EPIC 211966629 in Praesepe. Their chromospheric emission remains relatively constant as a function of rotational phase. There is one exceptional rotation of EPIC 211928486 where the $S$-index declined by 50\%, without the expected corresponding increase in broadband stellar flux, for which we have few good explanations. Future simultaneous observations will be possible with NASA's TESS mission, which may allow us to determine whether the uncorrelated $S$-index dipping phenomenon occurs regularly for this star or mid K dwarfs like it.


\begin{thebibliography}{}
\expandafter\ifx\csname natexlab\endcsname\relax\def\natexlab#1{#1}\fi
\providecommand{\url}[1]{\href{#1}{#1}}

\bibitem[{{Aigrain} {et~al.}(2015){Aigrain}, {Llama}, {Ceillier}, {Chagas},
  {Davenport}, {Garc{\'{\i}}a}, {Hay}, {Lanza}, {McQuillan}, {Mazeh}, {de
  Medeiros}, {Nielsen}, \& {Reinhold}}]{Aigrain2015}
{Aigrain}, S., {Llama}, J., {Ceillier}, T., {et~al.} 2015, \mnras, 450, 3211

\bibitem[{{Allen} \& {Strom}(1995)}]{Allen1995}
{Allen}, L.~E., \& {Strom}, K.~M. 1995, \aj, 109, 1379

\bibitem[{{Andrae} {et~al.}(2018){Andrae}, {Fouesneau}, {Creevey}, {Ordenovic},
  {Mary}, {Burlacu}, {Chaoul}, {Jean-Antoine-Piccolo}, {Kordopatis}, {Korn},
  {Lebreton}, {Panem}, {Pichon}, {Thevenin}, {Walmsley}, \&
  {Bailer-Jones}}]{DR2prop}
{Andrae}, R., {Fouesneau}, M., {Creevey}, O., {et~al.} 2018, ArXiv e-prints,
  arXiv:1804.09374

\bibitem[{Bailyn(1995)}]{bluestraggler}
Bailyn, C.~D. 1995, Annual Review of Astronomy and Astrophysics, 33, 133.
\newblock \url{https://doi.org/10.1146/annurev.aa.33.090195.001025}

\bibitem[{{Bouvier} {et~al.}(2018){Bouvier}, {Barrado}, {Moraux}, {Stauffer},
  {Rebull}, {Hillenbrand}, {Bayo}, {Boisse}, {Bouy}, {DiFolco}, {Lillo-Box}, \&
  {Calder{\'o}n}}]{Bouvier2018}
{Bouvier}, J., {Barrado}, D., {Moraux}, E., {et~al.} 2018, \aap, 613, A63

\bibitem[{{Boyajian} {et~al.}(2012){Boyajian}, {von Braun}, {van Belle},
  {McAlister}, {ten Brummelaar}, {Kane}, {Muirhead}, {Jones}, {White},
  {Schaefer}, {Ciardi}, {Henry}, {L{\'o}pez-Morales}, {Ridgway}, {Gies}, {Jao},
  {Rojas-Ayala}, {Parks}, {Sturmann}, {Sturmann}, {Turner}, {Farrington},
  {Goldfinger}, \& {Berger}}]{Boyajian2012}
{Boyajian}, T.~S., {von Braun}, K., {van Belle}, G., {et~al.} 2012, \apj, 757,
  112

\bibitem[{{Brandt} \& {Huang}(2015)}]{Brandt2015}
{Brandt}, T.~D., \& {Huang}, C.~X. 2015, \apj, 807, 24

\bibitem[{{Bressan} {et~al.}(2012){Bressan}, {Marigo}, {Girardi}, {Salasnich},
  {Dal Cero}, {Rubele}, \& {Nanni}}]{parsec}
{Bressan}, A., {Marigo}, P., {Girardi}, L., {et~al.} 2012, ArXiv e-prints,
  arXiv:1208.4498

\bibitem[{{Brewer} {et~al.}(2015){Brewer}, {Fischer}, {Basu}, {Valenti}, \&
  {Piskunov}}]{Brewer2015}
{Brewer}, J.~M., {Fischer}, D.~A., {Basu}, S., {Valenti}, J.~A., \& {Piskunov},
  N. 2015, \apj, 805, 126

\bibitem[{{Brewer} {et~al.}(2016){Brewer}, {Fischer}, {Valenti}, \&
  {Piskunov}}]{Brewer2016}
{Brewer}, J.~M., {Fischer}, D.~A., {Valenti}, J.~A., \& {Piskunov}, N. 2016,
  \apjs, 225, 32

\bibitem[{{Carrington}(1859)}]{Carrington1859}
{Carrington}, R.~C. 1859, \mnras, 20, 13

\bibitem[{{Choi} {et~al.}(2016){Choi}, {Dotter}, {Conroy}, {Cantiello},
  {Paxton}, \& {Johnson}}]{Choi2016}
{Choi}, J., {Dotter}, A., {Conroy}, C., {et~al.} 2016, \apj, 823, 102

\bibitem[{{Cummings} {et~al.}(2017){Cummings}, {Deliyannis}, {Maderak}, \&
  {Steinhauer}}]{Cummings2017}
{Cummings}, J.~D., {Deliyannis}, C.~P., {Maderak}, R.~M., \& {Steinhauer}, A.
  2017, \aj, 153, 128

\bibitem[{{Davenport} {et~al.}(2015){Davenport}, {Hebb}, \&
  {Hawley}}]{Davenport2015}
{Davenport}, J.~R.~A., {Hebb}, L., \& {Hawley}, S.~L. 2015, \apj, 806, 212

\bibitem[{{Deming} {et~al.}(2011){Deming}, {Sada}, {Jackson}, {Peterson},
  {Agol}, {Knutson}, {Jennings}, {Haase}, \& {Bays}}]{Deming2011}
{Deming}, D., {Sada}, P.~V., {Jackson}, B., {et~al.} 2011, \apj, 740, 33

\bibitem[{{Dorren} \& {Guinan}(1982)}]{Dorren1982}
{Dorren}, J.~D., \& {Guinan}, E.~F. 1982, \aj, 87, 1546

\bibitem[{{Douglas} {et~al.}(2016){Douglas}, {Ag{\"u}eros}, {Covey}, {Cargile},
  {Barclay}, {Cody}, {Howell}, \& {Kopytova}}]{Douglas2016}
{Douglas}, S.~T., {Ag{\"u}eros}, M.~A., {Covey}, K.~R., {et~al.} 2016, \apj,
  822, 47

\bibitem[{{Douglas} {et~al.}(2017){Douglas}, {Ag{\"u}eros}, {Covey}, \&
  {Kraus}}]{Douglas2017}
{Douglas}, S.~T., {Ag{\"u}eros}, M.~A., {Covey}, K.~R., \& {Kraus}, A. 2017,
  \apj, 842, 83

\bibitem[{{Douglas} {et~al.}(2014){Douglas}, {Ag{\"u}eros}, {Covey}, {Bowsher},
  {Bochanski}, {Cargile}, {Kraus}, {Law}, {Lemonias}, {Arce}, {Fierroz}, \&
  {Kundert}}]{Douglas2014}
{Douglas}, S.~T., {Ag{\"u}eros}, M.~A., {Covey}, K.~R., {et~al.} 2014, \apj,
  795, 161

\bibitem[{{Eaton} \& {Hall}(1979)}]{Eaton1979}
{Eaton}, J.~A., \& {Hall}, D.~S. 1979, \apj, 227, 907

\bibitem[{{Foreman-Mackey} {et~al.}(2013){Foreman-Mackey}, {Hogg}, {Lang}, \&
  {Goodman}}]{Foreman-Mackey2013}
{Foreman-Mackey}, D., {Hogg}, D.~W., {Lang}, D., \& {Goodman}, J. 2013, \pasp,
  125, 306

\bibitem[{{Gaia Collaboration} {et~al.}(2018{\natexlab{a}}){Gaia
  Collaboration}, {Brown}, {Vallenari}, {Prusti}, {de Bruijne}, {Babusiaux}, \&
  {Bailer-Jones}}]{GaiaDR2}
{Gaia Collaboration}, {Brown}, A.~G.~A., {Vallenari}, A., {et~al.}
  2018{\natexlab{a}}, ArXiv e-prints, arXiv:1804.09365

\bibitem[{{Gaia Collaboration} {et~al.}(2018{\natexlab{b}}){Gaia
  Collaboration}, {Babusiaux}, {van Leeuwen}, {Barstow}, {Jordi}, {Vallenari},
  {Bossini}, {Bressan}, {Cantat-Gaudin}, {van Leeuwen}, \& et~al.}]{DR2HRD}
{Gaia Collaboration}, {Babusiaux}, C., {van Leeuwen}, F., {et~al.}
  2018{\natexlab{b}}, ArXiv e-prints, arXiv:1804.09378

\bibitem[{{Gossage} {et~al.}(2018){Gossage}, {Conroy}, {Dotter}, {Choi},
  {Rosenfield}, {Cargile}, \& {Dolphin}}]{Gossage2018}
{Gossage}, S., {Conroy}, C., {Dotter}, A., {et~al.} 2018, ArXiv e-prints,
  arXiv:1804.06441

\bibitem[{{Gu}(2005)}]{Gu2005}
{Gu}, S.-H. 2005, in IAU Symposium, Vol. 226, Coronal and Stellar Mass
  Ejections, ed. K.~{Dere}, J.~{Wang}, \& Y.~{Yan}, 501--505

\bibitem[{{Guglielmino} {et~al.}(2010){Guglielmino}, {Bellot Rubio},
  {Zuccarello}, {Aulanier}, {Vargas Dom{\'{\i}}nguez}, \&
  {Kamio}}]{Guglielmino2010}
{Guglielmino}, S.~L., {Bellot Rubio}, L.~R., {Zuccarello}, F., {et~al.} 2010,
  \apj, 724, 1083

\bibitem[{{Hall}(2008)}]{Hall2008}
{Hall}, J.~C. 2008, Living Reviews in Solar Physics, 5, 2

\bibitem[{{Honda} {et~al.}(2015){Honda}, {Notsu}, {Maehara}, {Notsu},
  {Shibayama}, {Nogami}, \& {Shibata}}]{Honda2015}
{Honda}, S., {Notsu}, Y., {Maehara}, H., {et~al.} 2015, \pasj, 67, 85

\bibitem[{{Howard} {et~al.}(1984){Howard}, {Gilman}, \& {Gilman}}]{Howard1984}
{Howard}, R., {Gilman}, P.~I., \& {Gilman}, P.~A. 1984, \apj, 283, 373

\bibitem[{{Howell} {et~al.}(2014){Howell}, {Sobeck}, {Haas}, {Still},
  {Barclay}, {Mullally}, {Troeltzsch}, {Aigrain}, {Bryson}, {Caldwell},
  {Chaplin}, {Cochran}, {Huber}, {Marcy}, {Miglio}, {Najita}, {Smith},
  {Twicken}, \& {Fortney}}]{k2}
{Howell}, S.~B., {Sobeck}, C., {Haas}, M., {et~al.} 2014, \pasp, 126, 398

\bibitem[{{Hunter}(2007)}]{matplotlib}
{Hunter}, J.~D. 2007, Computing in Science and Engineering, 9, 90

\bibitem[{{Husser} {et~al.}(2013){Husser}, {Wende-von Berg}, {Dreizler},
  {Homeier}, {Reiners}, {Barman}, \& {Hauschildt}}]{Husser2013}
{Husser}, T.-O., {Wende-von Berg}, S., {Dreizler}, S., {et~al.} 2013, \aap,
  553, A6

\bibitem[{{Isaacson} \& {Fischer}(2010)}]{Isaacson2010}
{Isaacson}, H., \& {Fischer}, D. 2010, \apj, 725, 875

\bibitem[{Jones {et~al.}(2001)Jones, Oliphant, Peterson, {et~al.}}]{scipy}
Jones, E., Oliphant, T., Peterson, P., {et~al.} 2001, {SciPy}: Open source
  scientific tools for {Python}, , .
\newblock \url{http://www.scipy.org/}

\bibitem[{{Katz} {et~al.}(2018){Katz}, {Sartoretti}, {Cropper}, {Panuzzo},
  {Seabroke}, {Viala}, {Benson}, {Blomme}, {Jasniewicz}, {Jean-Antoine},
  {Huckle}, {Smith}, {Baker}, {Crifo}, {Damerdji}, {David}, {Dolding},
  {Fr{\'e}mat}, {Gosset}, {Guerrier}, {Guy}, {Haigron}, {Jan{\ss}en},
  {Marchal}, {Plum}, {Soubiran}, {Th{\'e}venin}, {Ajaj}, {Allende Prieto},
  {Babusiaux}, {Boudreault}, {Chemin}, {Delle Luche}, {Fabre}, {Gueguen},
  {Hambly}, {Lasne}, {Meynadier}, {Pailler}, {Panem}, {Royer}, {Tauran},
  {Zurbach}, {Zwitter}, {Arenou}, {Bossini}, {Gomez}, {Lemaitre}, {Leclerc},
  {Morel}, {Munari}, {Turon}, {Vallenari}, \& {{\v Z}erjal}}]{DR2RV1}
{Katz}, D., {Sartoretti}, P., {Cropper}, M., {et~al.} 2018, ArXiv e-prints,
  arXiv:1804.09372

\bibitem[{{Kowalski} {et~al.}(2013){Kowalski}, {Hawley}, {Wisniewski}, {Osten},
  {Hilton}, {Holtzman}, {Schmidt}, \& {Davenport}}]{Kowalski2013}
{Kowalski}, A.~F., {Hawley}, S.~L., {Wisniewski}, J.~P., {et~al.} 2013, The
  Astrophysical Journal Supplement Series, 207, 15

\bibitem[{{Lockwood} {et~al.}(1984){Lockwood}, {Thompson}, {Radick}, {Osborn},
  {Baggett}, {Duncan}, \& {Hartmann}}]{Lockwood1984}
{Lockwood}, G.~W., {Thompson}, D.~T., {Radick}, R.~R., {et~al.} 1984, \pasp,
  96, 714

\bibitem[{{Luo} {et~al.}(2015){Luo}, {Zhao}, {Zhao}, {Deng}, {Liu}, {Jing},
  {Wang}, {Zhang}, {Shi}, {Cui}, {Chu}, {Li}, {Bai}, {Wu}, {Cai}, {Cao}, {Cao},
  {Carlin}, {Chen}, {Chen}, {Chen}, {Chen}, {Chen}, {Chen}, {Chen},
  {Christlieb}, {Chu}, {Cui}, {Dong}, {Du}, {Fan}, {Feng}, {Fu}, {Gao}, {Gong},
  {Gu}, {Guo}, {Han}, {He}, {Hou}, {Hou}, {Hou}, {Hu}, {Hu}, {Hu}, {Huo},
  {Jia}, {Jiang}, {Jiang}, {Jiang}, {Jin}, {Kong}, {Kong}, {Lei}, {Li}, {Li},
  {Li}, {Li}, {Li}, {Li}, {Li}, {Li}, {Li}, {Li}, {Li}, {Li}, {Liang}, {Lin},
  {Liu}, {Liu}, {Liu}, {Liu}, {Lu}, {Luo}, {Mao}, {Newberg}, {Ni}, {Qi}, {Qi},
  {Shen}, {Shi}, {Song}, {Song}, {Su}, {Su}, {Tang}, {Tao}, {Tian}, {Wang},
  {Wang}, {Wang}, {Wang}, {Wang}, {Wang}, {Wang}, {Wang}, {Wang}, {Wang},
  {Wang}, {Wang}, {Wang}, {Wang}, {Wang}, {Wang}, {Wang}, {Wang}, {Wang},
  {Wang}, {Wei}, {Wei}, {Wu}, {Wu}, {Wu}, {Wu}, {Xing}, {Xu}, {Xu}, {Xu},
  {Yan}, {Yang}, {Yang}, {Yang}, {Yang}, {Yao}, {Yu}, {Yuan}, {Yuan}, {Yuan},
  {Yuan}, {Zhai}, {Zhang}, {Zhang}, {Zhang}, {Zhang}, {Zhang}, {Zhang},
  {Zhang}, {Zhang}, {Zhao}, {Zhou}, {Zhou}, {Zhu}, {Zhu}, {Zou}, \&
  {Zuo}}]{Luo2015}
{Luo}, A.-L., {Zhao}, Y.-H., {Zhao}, G., {et~al.} 2015, Research in Astronomy
  and Astrophysics, 15, 1095

\bibitem[{{Mamajek} \& {Hillenbrand}(2008)}]{Mamajek2008}
{Mamajek}, E.~E., \& {Hillenbrand}, L.~A. 2008, \apj, 687, 1264

\bibitem[{{Mandal} {et~al.}(2017){Mandal}, {Chatterjee}, \&
  {Banerjee}}]{Mandal2017}
{Mandal}, S., {Chatterjee}, S., \& {Banerjee}, D. 2017, \apj, 835, 158

\bibitem[{{Marigo} {et~al.}(2017){Marigo}, {Girardi}, {Bressan}, {Rosenfield},
  {Aringer}, {Chen}, {Dussin}, {Nanni}, {Pastorelli}, {Rodrigues}, {Trabucchi},
  {Bladh}, {Dalcanton}, {Groenewegen}, {Montalb{\'a}n}, \& {Wood}}]{Marigo2017}
{Marigo}, P., {Girardi}, L., {Bressan}, A., {et~al.} 2017, \apj, 835, 77

\bibitem[{{McQuillan} {et~al.}(2013){McQuillan}, {Mazeh}, \&
  {Aigrain}}]{McQuillan2013}
{McQuillan}, A., {Mazeh}, T., \& {Aigrain}, S. 2013, \apj, 775, L11

\bibitem[{{McQuillan} {et~al.}(2014){McQuillan}, {Mazeh}, \&
  {Aigrain}}]{McQuillan2014}
---. 2014, \apjs, 211, 24

\bibitem[{Morris(2017)}]{arceshk}
Morris, B.~M. 2017, bmorris3/arces\_hk: v0.1, , , doi:10.5281/zenodo.886630.
\newblock \url{https://doi.org/10.5281/zenodo.886630}

\bibitem[{Morris \& Dorn-Wallenstein(2018)}]{aesop}
Morris, B.~M., \& Dorn-Wallenstein, T. 2018, Journal of Open Source Software,
  3, 854.
\newblock \url{https://doi.org/10.21105/joss.00854}

\bibitem[{{Morris} {et~al.}(2018{\natexlab{a}}){Morris}, {Hawley}, \&
  {Hebb}}]{Morris2018d}
{Morris}, B.~M., {Hawley}, S.~L., \& {Hebb}, L. 2018{\natexlab{a}}, Research
  Notes of the American Astronomical Society, 2, 26

\bibitem[{{Morris} {et~al.}(2017{\natexlab{a}}){Morris}, {Hebb}, {Davenport},
  {Rohn}, \& {Hawley}}]{Morris2017a}
{Morris}, B.~M., {Hebb}, L., {Davenport}, J.~R.~A., {Rohn}, G., \& {Hawley},
  S.~L. 2017{\natexlab{a}}, \apj, 846, 99

\bibitem[{{Morris} {et~al.}(2017{\natexlab{b}}){Morris}, {Hawley}, {Hebb},
  {Sakari}, {Davenport}, {Isaacson}, {Howard}, {Montet}, \&
  {Agol}}]{Morris2017b}
{Morris}, B.~M., {Hawley}, S.~L., {Hebb}, L., {et~al.} 2017{\natexlab{b}},
  \apj, 848, 58

\bibitem[{{Morris} {et~al.}(2018{\natexlab{b}}){Morris}, {Tollerud}, {Sip{\H
  o}cz}, {Deil}, {Douglas}, {Berlanga Medina}, {Vyhmeister}, {Smith},
  {Littlefair}, {Price-Whelan}, {Gee}, \& {Jeschke}}]{astroplan}
{Morris}, B.~M., {Tollerud}, E., {Sip{\H o}cz}, B., {et~al.}
  2018{\natexlab{b}}, \aj, 155, 128

\bibitem[{{Neuh{\"a}user} \& {Hambaryan}(2014)}]{Neuhauser2014}
{Neuh{\"a}user}, R., \& {Hambaryan}, V.~V. 2014, Astronomische Nachrichten,
  335, 949

\bibitem[{{Notsu} {et~al.}(2015{\natexlab{a}}){Notsu}, {Honda}, {Maehara},
  {Notsu}, {Shibayama}, {Nogami}, \& {Shibata}}]{Notsu2015a}
{Notsu}, Y., {Honda}, S., {Maehara}, H., {et~al.} 2015{\natexlab{a}}, \pasj,
  67, 32

\bibitem[{{Notsu} {et~al.}(2015{\natexlab{b}}){Notsu}, {Honda}, {Maehara},
  {Notsu}, {Shibayama}, {Nogami}, \& {Shibata}}]{Notsu2015b}
---. 2015{\natexlab{b}}, \pasj, 67, 33

\bibitem[{{O'Neal} {et~al.}(2004){O'Neal}, {Neff}, {Saar}, \&
  {Cuntz}}]{ONeal2004}
{O'Neal}, D., {Neff}, J.~E., {Saar}, S.~H., \& {Cuntz}, M. 2004, \aj, 128, 1802

\bibitem[{P\'erez \& Granger(2007)}]{ipython}
P\'erez, F., \& Granger, B.~E. 2007, Computing in Science and Engineering, 9,
  21.
\newblock \url{http://ipython.org}

\bibitem[{{Piskunov} \& {Valenti}(2017)}]{Piskunov2017}
{Piskunov}, N., \& {Valenti}, J.~A. 2017, \aap, 597, A16

\bibitem[{{Radick} {et~al.}(2018){Radick}, {Lockwood}, {Henry}, {Hall}, \&
  {Pevtsov}}]{Radick2018}
{Radick}, R.~R., {Lockwood}, G.~W., {Henry}, G.~W., {Hall}, J.~C., \&
  {Pevtsov}, A.~A. 2018, \apj, 855, 75

\bibitem[{{Radick} {et~al.}(1998){Radick}, {Lockwood}, {Skiff}, \&
  {Baliunas}}]{Radick1998}
{Radick}, R.~R., {Lockwood}, G.~W., {Skiff}, B.~A., \& {Baliunas}, S.~L. 1998,
  \apjs, 118, 239

\bibitem[{{Radick} {et~al.}(1983){Radick}, {Wilkerson}, {Worden}, {Africano},
  {Klimke}, {Ruden}, {Rogers}, {Armandroff}, \& {Giampapa}}]{Radick1983}
{Radick}, R.~R., {Wilkerson}, M.~S., {Worden}, S.~P., {et~al.} 1983, \pasp, 95,
  300

\bibitem[{{Sanchis-Ojeda} \& {Winn}(2011)}]{Sanchis-Ojeda2011}
{Sanchis-Ojeda}, R., \& {Winn}, J.~N. 2011, \apj, 743, 61

\bibitem[{{Shapiro} {et~al.}(2016){Shapiro}, {Solanki}, {Krivova}, {Yeo}, \&
  {Schmutz}}]{Shapiro2016}
{Shapiro}, A.~I., {Solanki}, S.~K., {Krivova}, N.~A., {Yeo}, K.~L., \&
  {Schmutz}, W.~K. 2016, \aap, 589, A46

\bibitem[{{Solanki}(2003)}]{Solanki2003}
{Solanki}, S.~K. 2003, \aapr, 11, 153

\bibitem[{{Soubiran} {et~al.}(2018){Soubiran}, {Jasniewicz}, {Chemin},
  {Zurbach}, {Brouillet}, {Panuzzo}, {Sartoretti}, {Katz}, {Le Campion},
  {Marchal}, {Hestroffer}, {Th{\'e}venin}, {Crifo}, {Udry}, {Cropper},
  {Seabroke}, {Viala}, {Benson}, {Blomme}, {Jean-Antoine}, {Huckle}, {Smith},
  {Baker}, {Damerdji}, {Dolding}, {Fr{\'e}mat}, {Gosset}, {Guerrier}, {Guy},
  {Haigron}, {Jan{\ss}en}, {Plum}, {Fabre}, {Lasne}, {Pailler}, {Panem},
  {Riclet}, {Royer}, {Tauran}, {Zwitter}, {Gueguen}, \& {Turon}}]{DR2RV2}
{Soubiran}, C., {Jasniewicz}, G., {Chemin}, L., {et~al.} 2018, ArXiv e-prints,
  arXiv:1804.09370

\bibitem[{{Spruit}(1977)}]{Spruit1977}
{Spruit}, H.~C. 1977, \solphys, 55, 3

\bibitem[{{Stauffer} {et~al.}(1998){Stauffer}, {Schultz}, \&
  {Kirkpatrick}}]{Stauffer1998}
{Stauffer}, J.~R., {Schultz}, G., \& {Kirkpatrick}, J.~D. 1998, \apjl, 499,
  L199

\bibitem[{{Stix}(1989)}]{Stix1989}
{Stix}, M. 1989, {The Sun. an Introduction}, 192

\bibitem[{{Taylor}(2006)}]{Taylor2006}
{Taylor}, B.~J. 2006, \aj, 132, 2453

\bibitem[{{The Astropy Collaboration} {et~al.}(2018){The Astropy
  Collaboration}, {Price-Whelan}, {Sip{\H o}cz}, {G{\"u}nther}, {Lim},
  {Crawford}, {Conseil}, {Shupe}, {Craig}, {Dencheva}, {Ginsburg},
  {VanderPlas}, {Bradley}, {P{\'e}rez-Su{\'a}rez}, {de Val-Borro}, {Aldcroft},
  {Cruz}, {Robitaille}, {Tollerud}, {Ardelean}, {Babej}, {Bachetti}, {Bakanov},
  {Bamford}, {Barentsen}, {Barmby}, {Baumbach}, {Berry}, {Biscani}, {Boquien},
  {Bostroem}, {Bouma}, {Brammer}, {Bray}, {Breytenbach}, {Buddelmeijer},
  {Burke}, {Calderone}, {Cano Rodr{\'{\i}}guez}, {Cara}, {Cardoso},
  {Cheedella}, {Copin}, {Crichton}, {D{\'A}vella}, {Deil}, {Depagne},
  {Dietrich}, {Donath}, {Droettboom}, {Earl}, {Erben}, {Fabbro}, {Ferreira},
  {Finethy}, {Fox}, {Garrison}, {Gibbons}, {Goldstein}, {Gommers}, {Greco},
  {Greenfield}, {Groener}, {Grollier}, {Hagen}, {Hirst}, {Homeier}, {Horton},
  {Hosseinzadeh}, {Hu}, {Hunkeler}, {Ivezi{\'c}}, {Jain}, {Jenness}, {Kanarek},
  {Kendrew}, {Kern}, {Kerzendorf}, {Khvalko}, {King}, {Kirkby}, {Kulkarni},
  {Kumar}, {Lee}, {Lenz}, {Littlefair}, {Ma}, {Macleod}, {Mastropietro},
  {McCully}, {Montagnac}, {Morris}, {Mueller}, {Mumford}, {Muna}, {Murphy},
  {Nelson}, {Nguyen}, {Ninan}, {N{\"o}the}, {Ogaz}, {Oh}, {Parejko}, {Parley},
  {Pascual}, {Patil}, {Patil}, {Plunkett}, {Prochaska}, {Rastogi}, {Reddy
  Janga}, {Sabater}, {Sakurikar}, {Seifert}, {Sherbert}, {Sherwood-Taylor},
  {Shih}, {Sick}, {Silbiger}, {Singanamalla}, {Singer}, {Sladen}, {Sooley},
  {Sornarajah}, {Streicher}, {Teuben}, {Thomas}, {Tremblay}, {Turner},
  {Terr{\'o}n}, {van Kerkwijk}, {de la Vega}, {Watkins}, {Weaver}, {Whitmore},
  {Woillez}, \& {Zabalza}}]{Astropy2018}
{The Astropy Collaboration}, {Price-Whelan}, A.~M., {Sip{\H o}cz}, B.~M.,
  {et~al.} 2018, ArXiv e-prints, arXiv:1801.02634

\bibitem[{{Valenti} \& {Fischer}(2005)}]{Valenti2005}
{Valenti}, J.~A., \& {Fischer}, D.~A. 2005, \apjs, 159, 141

\bibitem[{{Valenti} \& {Piskunov}(1996)}]{sme}
{Valenti}, J.~A., \& {Piskunov}, N. 1996, \aaps, 118, 595

\bibitem[{{Van Der Walt} {et~al.}(2011){Van Der Walt}, {Colbert}, \&
  {Varoquaux}}]{VanDerWalt2011}
{Van Der Walt}, S., {Colbert}, S.~C., \& {Varoquaux}, G. 2011, ArXiv e-prints,
  arXiv:1102.1523

\bibitem[{Vanderplas(2015)}]{gatspy}
Vanderplas, J. 2015, {gatspy: General tools for Astronomical Time Series in
  Python}, , , doi:10.5281/zenodo.14833.
\newblock \url{http://dx.doi.org/10.5281/zenodo.14833}

\bibitem[{{VanderPlas} \& {Ivezi{\'c}}(2015)}]{VanderPlas2015}
{VanderPlas}, J.~T., \& {Ivezi{\'c}}, {\v{Z}}. 2015, \apj, 812, 18

\bibitem[{{Walkowicz} \& {Basri}(2013)}]{Walkowicz2013}
{Walkowicz}, L.~M., \& {Basri}, G.~S. 2013, \mnras, 436, 1883

\bibitem[{{Wright} {et~al.}(2004){Wright}, {Marcy}, {Butler}, \&
  {Vogt}}]{Wright2004}
{Wright}, J.~T., {Marcy}, G.~W., {Butler}, R.~P., \& {Vogt}, S.~S. 2004, \apjs,
  152, 261

\end{thebibliography}

\acknowledgements

We gratefully acknowledge support from NSF grant AST-1312453. 
S.T.D.~acknowledges support provided by the NSF through grant AST-1701468.

Based on observations obtained with the Apache Point Observatory 3.5-meter telescope, which is owned and operated by the Astrophysical Research Consortium. IRAF is distributed by the National Optical Astronomy Observatory, which is operated by the Association of Universities for Research in Astronomy, Inc., under cooperative agreement with the National Science Foundation. This paper includes data collected by the K2 mission. Funding for the K2 mission is provided by the NASA Science Mission directorate. This work presents results from the European Space Agency (ESA) space mission Gaia. Gaia data are being processed by the Gaia Data Processing and Analysis Consortium (DPAC). Funding for the DPAC is provided by national institutions, in particular the institutions participating in the Gaia MultiLateral Agreement (MLA). The Gaia mission website is \url{https://www.cosmos.esa.int/gaia}. The Gaia archive website is \url{https://archives.esac.esa.int/gaia}. Guoshoujing Telescope (the Large Sky Area Multi-Object Fiber Spectroscopic Telescope, or LAMOST) is a National Major Scientific Project built by the Chinese Academy of Sciences. Funding for the project has been provided by the National Development and Reform Commission. LAMOST is operated and managed by the National Astronomical Observatories, Chinese Academy of Sciences. Some of the data presented in this paper were obtained from the Mikulski Archive for Space Telescopes (MAST). STScI is operated by the Association of Universities for Research in Astronomy, Inc., under NASA contract NAS5-26555.

\facilities{K2, APO/ARC 3.5 m, APO/ARCSAT} 

\software{\texttt{astroplan} \citep{astroplan}, \texttt{astropy} \citep{Astropy2018}, \texttt{arces\_hk} \citep{arceshk}, \texttt{gatspy} \citep{gatspy}, \texttt{emcee} \citep{Foreman-Mackey2013}, \texttt{ipython} \citep{ipython}, \texttt{numpy} \citep{VanDerWalt2011}, \texttt{scipy} \citep{scipy},  \texttt{matplotlib} \citep{matplotlib}, \texttt{SME} \citep{sme}}

\end{document}